\documentclass[twocolumn,english,prb]{revtex4}
\usepackage{pslatex}
\usepackage[T1]{fontenc}
\usepackage[latin1]{inputenc}
\usepackage{graphicx}

\makeatletter

\usepackage{babel}
\makeatother
\begin{document}

\title{X-ray method to study temperature-dependent stripe domains in MnAs/GaAs(001)}

\author{R. Magalhães-Paniago$^{a,b}$, L.N. Coelho$^{a}$, B.R.A. Neves$^{a}$,
\\
H. Westfahl Jr.$^{b}$, F. Iikawa$^{c}$, L. Däweritz$^{d}$, C. Spezzani$^{e}$,
and M. Sacchi$^{e}$}

\affiliation{$^{a}$Departamento de Física, Universidade Federal de Minas Gerais,
CP 702, Belo Horizonte MG, 30123-970 Brazil.\\
$^{b}$Laboratório Nacional de Luz Síncrotron, CP 6192, Campinas SP,
13084-971, Brazil. \\
$^{c}$Instituto de Física \char`\"{}Gleb Wataghin\char`\"{}, UNICAMP,
CP 6165, 13083-970 SP, Brazil.\\
$^{d}$Paul-Drude-Institut für Festkörperelektronik, Hausvogteiplatz
5-7, 10117 Berlin, Germany.\\
$^{e}$Laboratoire pour l'Utilisation du Rayonnement Electromagnétique,
Centre Universitaire Paris-Sud, Boîte Postale 34, 91898 Orsay, France.}

\begin{abstract}
MnAs films grown on GaAs (001) exhibit a progressive transition between
hexagonal (ferromagnetic) and orthorhombic (paramagnetic) phases at
wide temperature range instead of abrupt transition during the first-order
phase transition. The coexistence of two phases is favored by the
anisotropic strain arising from the constraint on the MnAs films imposed
by the substrate. This phase coexistence occurs in ordered arrangement
alternating periodic terrace steps. We present here a method to study
the surface morphology throughout this transition by means of specular
and diffuse scattering of soft x-rays, tuning the photon energy at
the Mn 2p resonance. The results show the long-range arrangement of
the periodic stripe-like structure during the phase coexistence and
its period remains constant, in agreement with previous results using
other techniques.
\end{abstract}
\maketitle
The integration of magnetic films in semiconductor devices represents
one of the major challenges in materials science. MnAs is a very promising
material for spin injection, in combination with III-V semiconductor
compounds (such as GaAs) \cite{1,2}. Around room temperature (RT),
bulk MnAs undergoes a first order structural and magnetic phase transition
\cite{key-3,key-4,key-5} between the low temperature $\alpha$ phase,
hexagonal and ferromagnetic, and the high temperature $\beta$ phase,
orthorhombic and paramagnetic. MnAs films grown on GaAs(001) \cite{key-6,key-7,key-8}
exhibit a phase coexistence over a temperature range that can extend
between 0 °C and 55 °C, depending on the thickness of the film \cite{key-9,key-10,key-11,key-12}.
Several local probe techniques have been used to characterize the
surface morphology and magnetic structure, such as atomic force microscopy
(AFM) \cite{key-12}, magnetic force microscopy (MFM) \cite{key-13}
and X-ray magnetic circular dichroism photo-emission electron microscopy
(XMCDPEEM) \cite{key-14}. They highlighted the formation of periodic
terraces and magnetic domains at intermediate temperatures for the
phase transition. Information on the long range order of terraces
is usually gained through scattering experiments. In the case of MnAs,
x-ray diffraction experiments showed the coexistence of the two crystallographic
phases \cite{key-9,key-11}, but could not address the formation of
periodic domains with long range order.

In this work we have used specular and diffuse x-ray reflectivity
to follow the formation and evolution with temperature of the terrace
morphology throughout the $\alpha-\beta$ phase transition. In contrast
to the microscopy technique, such as atomic force microscopy, which
probes small area and only near the surface, the x-ray reflectivity
technique proposed here analyzes over a much larger area of the surface
and deepness, providing long range order information and also the
interface. On MnAs films analyzed using this technique we observe
the formation and disappearance of diffuse satellite peaks associated
with the terrace periodicity as a function of temperature. The modeling
of their intensity profiles yields numbers for the temperature-dependent
terrace widths.

A 130 nm thick MnAs was grown by molecular beam epitaxy on a GaAs(001)
substrate at 250 °C in a growth condition to obtain A-type orientation
\cite{key-8}, i.e., the MnAs plane in the $\alpha$ phase parallel
to the GaAs $\left(001\right)$ plane with the MnAs $\left[0001\right]$
c-axis along the GaAs direction.

We have first used AFM to identify the terrace-like steps discussed
before. Figure1a depicts a room temperature AFM image of the surface
topography of the MnAs film, which shows the formation of stripe-like
domains, elongated along the $\left(0001\right)$ direction, during
the coexistence of two phases and a clear alternate periodic structure
between $\alpha$ and $\beta$ phases is observed. Over the extension
of the AFM image, these terrace steps appear periodic, with a modulation
period of about 600 nm. The lower terraces correspond to regions where
no magnetic signal is obtained with Magnetic Force Microscopy (MFM),
therefore they can be associated to the high temperature paramagnetic
$\beta$ phase. The higher terraces (ferromagnetic $\alpha$ phase)
exhibit a complex magnetic structure where domains of opposite orientation
seem to be intercalated \cite{key-13,key-14}. Several authors have
already addressed the magnetic domain structure of these films using
MFM techniques, and at least three different types of domain formations
were observed \cite{key-13,key-14}. Here we will concentrate on our
X-ray method to study the terrace configuration.

In order to quantitatively evaluate the surface morphology of the
sample, we performed resonant x-ray scattering measurements at beamline
SU-7 of the SuperACO storage ring (LURE laboratory, Orsay). The beamline,
equipped with a linear undulator source, covers the 100-1000 eV range
with a resolving power of about 2000. The endstation is a 2-circle
($\omega/2\theta$) reflectometer working in ultra-high vacuum \cite{key-15}.
As shown in Fig.1b, the scattering geometry was co-planar, with the
incoming beam (of wave vector $\mathbf{k}_{i}$) impinging on 0.5mm
x 0.5mm of the MnAs film at a grazing angle , and the scattered photons
($\mathbf{k}_{f}$) collected at an angle $2\theta$ with respect
to the incident beam. The scattering vector $\mathbf{q}=\mathbf{k}_{f}-\mathbf{k}_{i}$
can be separated into two components, one parallel $q_{x}=\frac{2\pi}{\lambda}\left[\cos\left(2\theta-\omega\right)-\cos\left(\omega\right)\right]$
and one perpendicular $q_{z}=\frac{2\pi}{\lambda}\left[\sin\left(2\theta-\omega\right)+\sin\left(\omega\right)\right]$
to the sample surface. A structure that is periodic with a modulation
period $\ell$ will give rise to constructive interference in the
scattering process when $q=\frac{2\pi}{\ell}$.

\begin{figure}
\includegraphics[%
  width=1.0\columnwidth]{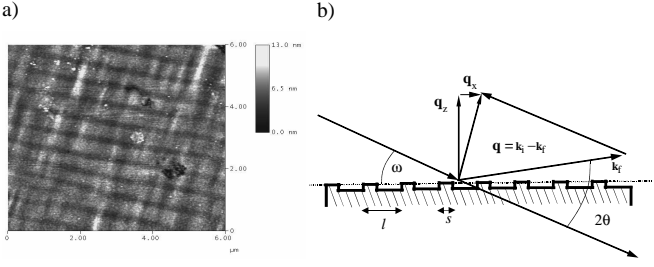}

\caption{\label{Fig1}{\scriptsize a) Atomic force microscopy image of the
130 nm thick MnAs film grown on GaAs(001); b) Scattering geometry
used to examine the terrace step structure. Scattering vectors are
described in the text.}}
\end{figure}

The horizontal terrace step structure can be studied in detail by
performing rocking scans of the sample, i.e. $\omega$ scans at fixed
$2\theta$. Over a limited range in $\omega$ around specular ($\omega=\frac{2\theta}{2}$),
a rocking scan corresponds to a $q_{x}$ scan at fixed $q_{z}$. In
Fig. 2a we show the map of the scattered intensity that results from
a series of 66 $q_{x}$ scans (201 points each) performed as a function
of temperature at a given $q_{z}$ value (1.68 nm-1, corresponding
to $2\theta=30^{\circ}$ and photon energy (640 eV). Fig. 2b shows
the line plots of three of these scans, at temperatures corresponding
to the $\alpha-\beta$ phase coexistence region. In the intermediate
temperature region, we observe, besides the specular peak (zero order)
at $q_{x}=0$, two other peaks (first order) at $q_{x}=\pm0.0107\, nm^{-1}$.
We ascribe them to the long range lateral order produced by the stripe
domains in the phase coexistence temperature region. The associated
modulation period is $\ell=\frac{2\pi}{q_{x}}=587\, nm$. It is worth
noticing that, while the intensity of the first order peaks varies
drastically as a function of temperature, their position remains unaltered.
This implies that the modulation period of the stripes, i.e. the sum
of the widths of the two domains with different structures, remains
constant. The $\alpha$ to $\beta$ transition takes place versus
temperature with the widths of the $\beta$ stripe increasing at the
expense of the width of the $\alpha$ phase, their sum remaining constant.
This has been previously observed by AFM \cite{key-12} and by XMCDPEEM
\cite{key-14}.

\begin{figure}
\includegraphics[%
  bb=15bp 15bp 304bp 221bp,
  scale=0.8]{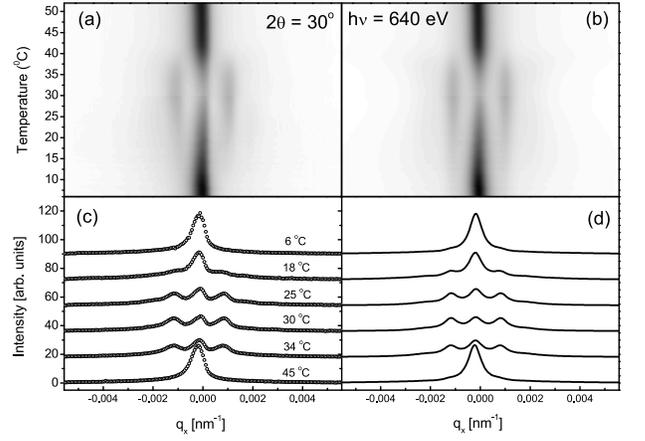}

\caption{\label{Fig2}{\scriptsize a) Two-dimensional plot of diffuse scattered
intensity measured at $2\theta=30^{\circ}$ as a function of $q_{x}$
(i.e. sample angle $\omega$) and of temperature in the phase coexistence
region and b) Corresponding fitted profiles following the model given
in the text; c) Selected X-ray scans for different temperatures; d)
Fitted profiles for corresponding scans on the left. Each profile
was fitted separately according to equation \ref{I4}.}}
\end{figure}

In order to reproduce the scattering profiles, a model was introduced.
Following Holy et al. \cite{key-16}, the scattering intensity stemming
from a lateral periodic structure is given by\[
I\left(q_{x},q_{z}\right)=\text{const.}\left|T_{i}T_{f}\right|^{2}\left|C\left(q_{x}\right)\right|^{^{2}}\left|F\left(q_{x},q_{z}\right)\right|^{2}\]
where $T_{i}$ and $T_{f}$ are the incident and exit transmission
functions, which depend on the incident and exit angles and can give
rise to multiple scattering for very thin films or multilayers (not
the present film). In this case one might need to address the behavior
of $T_{i}$ and $T_{f}$ with angle to understand the scattering data.

\[
F\left(q_{x},q_{z}\right)\equiv\int_{0}^{\ell}e^{-iq_{z}h\left(x\right)-iq_{x}x}dx\]
is the Fourier transform of the height profile function $h\left(x\right)$
of the surface, covering one terrace of each kind. In our case $h\left(x\right)$
is given by\[
h\left(x\right)\equiv\left\{ \begin{array}{c}
0\,,\,\textrm{if}\,\,\,0<x<s\\
h_{0}\,,\,\text{if}\,\, s<x<\ell\end{array}\right.\,,\]
 where $s$ is the width of the $\beta$ phase terrace and $\ell$
the sum of the widths of the two terraces. 

\[
C\left(q_{x}\right)\equiv\langle\sum_{m,n}\text{exp}\left[-i\left(R_{m}-R_{n}\right)q_{x}\right]\rangle\]
is the correlation function of different sets of two terraces averaged
over the whole sample surface. This correlation function is proportional
to a periodic sequence of $\delta$-like functions \cite{key-16}
centered at reciprocal lattice positions at intervals $g=\frac{2\pi}{\ell}$,
resulting in

\[
C\left(q_{x}\right)=N\langle g\sum_{p=-\infty}^{\infty}\delta\left(q_{x}-p\, g\right)\rangle,\]
where N is the number of periods. These $\delta$-like functions take
a Lorentzian-like lineshape with full-width $\sigma$. This finite
width here represents the combination of the limited coherence length
of the x-ray beam and the correlation length of the stripes. We obtain
the intensity profile, which in our was limited to the first 4 reciprocal
lattice points\begin{eqnarray}
I\left(q_{x},q_{z}\right) & \propto & \left|T_{i}T_{f}\right|^{2}\left\{ \frac{s^{2}+\left(\ell-s\right)^{2}+2\cos\left(q_{z}h\right)\ell\left(\ell-s\right)}{q_{x}^{2}+\sigma^{2}}\right.\label{I4}\\
 &  & \left.+16\sin\left(\frac{hq_{z}}{2}\right)\sum_{p=-4\left(p\neq0\right)}^{4}\frac{\frac{\sin\left(\frac{\pi ps}{\ell}\right)^{2}}{\left|p\right|^{3}\left(\frac{2\pi}{\ell}\right)^{2}}}{\left(\frac{q_{x}}{p}-\frac{2\pi}{\ell}\right)^{2}+\sigma^{2}}\right\} \nonumber \end{eqnarray}
In figures 2c and 2d the results of calculations are compared to the
corresponding experimental results of 2a and 2b. The intensity profiles
were fitted separately for each temperature, using $s$, $\ell$ and
$\sigma$ and an overall scaling factor as fitting parameters. This
model clearly reproduces extremely well the scattering data. The parameter
$\sigma=\left(3.2\pm0.1\right)\times10^{-4}nm^{-1}$ remained constant
throughout all measurements, which indicates that the major source
of broadening comes from the finite coherence length of the x-ray
beam.

In order to evaluate the relative weight of the $\alpha$ and $\beta$
phases, one can reduce the number of fitting parameters by limiting
the analysis to the ratio between the intensities of the first order
peak and of the specular reflectivity. In this case the intensity
ratio of these two peaks as predicted by equation \ref{I4} is approximately
given by \cite{key-17}

\begin{equation}
\frac{I\left(q_{x}=\frac{2\pi}{\ell}\right)}{I\left(q_{x}=0\right)}=\frac{16\sin\left(\frac{hq_{z}}{2}\right)\sin\left(\frac{\pi s}{\ell}\right)^{2}/\left(\frac{2\pi}{\ell}\right)^{2}}{s^{2}+\left(\ell-s\right)^{2}+2\cos\left(q_{z}h\right)\ell\left(\ell-s\right)}\label{IoverI}\end{equation}

The terrace width was determined by solving numerically equation \ref{IoverI}
independently from other fitting parameters. As a result, much smaller
error bars for the terrace width were obtained. Fig. 3 shows the temperature
dependence of the $\beta$-phase terrace width.

\begin{figure}
\includegraphics[%
  bb=15bp 15bp 282bp 213bp,
  clip,
  width=1.0\columnwidth]{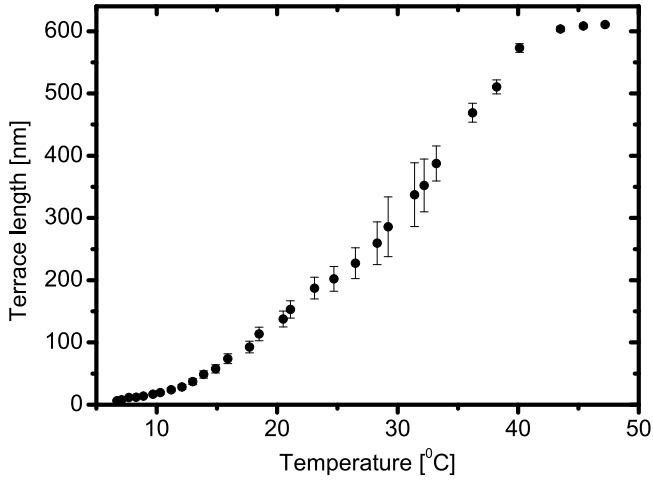}

\caption{{\scriptsize \label{Fig3}Temperature dependence of the $\beta$
phase terrace width. The error bars are estimated from the statistical
fluctuation of the scattered intensities of the specular and first
diffuse satellite peaks. The error bars are larger in the intermediate
temperature region due to the lower intensity of the specular peak.}}
\end{figure}

Finally, we have measured the scattered intensity as a function of
$q_{z}$ for a constant non-zero value of $q_{x}$. This can be obtained
by independently positioning sample and detector angles according
to a predetermined set of values. Fig. 4a shows the results of two
$q_{z}$ scans ($T=5^{\circ}C$ and $T=32^{\circ}C$) taken at $q_{x}=0.0107\, nm^{-1}$,
i.e. at $q_{x}$ value matching the lateral order parameter of the
stripes. Besides the peak of the transmission function (at $q_{z}=0.4\, nm^{-1}$)
present in scans perpendicular to the surface, at $T=32^{\circ}C$
there is a second peak centered around $q_{z}=1\, nm^{-1}$, which
comes from the interference between the waves scattered at the lower
and the higher terraces.

In order to eliminate the contribution of the transmission function
and to enhance the interference term, Fig. 4b shows the intensity
ratio of the two scans. This ratio is centered at $q_{z}=1.08\pm0.02\, nm^{-1}$,
corresponding to a difference in height between the two stripes of
$5.82\, nm$ (this height corresponds to an average value over the
whole sample surface).

\begin{figure}
\includegraphics[%
  scale=0.8]{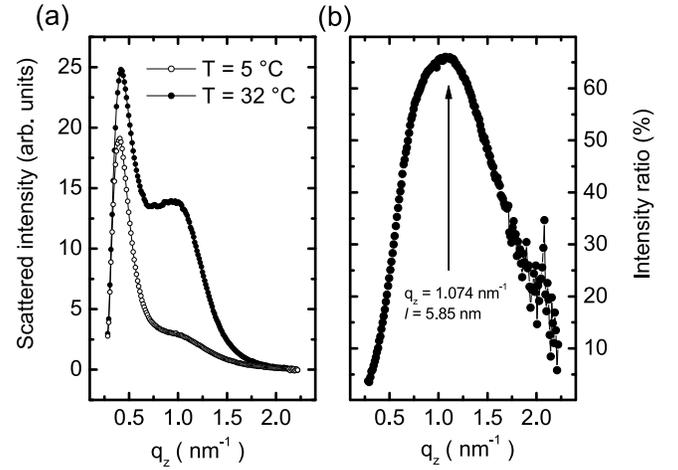}

\caption{{\scriptsize \label{Fig4}a) Scattered intensity as a function of
vertical momentum transfer $q_{z}$ for a parallel momentum transfer
$q_{z}$ for a parallel momentum transfer $q_{x}=0.0107\, nm^{-1}$.
Hollow circles: T = 5°C; filled circles: T = 32°C; b) Intensity ratio
$\frac{I\left(T=320C\right)-I\left(T=50C\right)}{I\left(T=320C\right)+I\left(T=50C\right)}$
of the two curves in a), used to determine the terrace height. }}
\end{figure}

In contrast to the surface morphologic microscopy techniques, the
X-ray technique used here probes wide area and the depth of the film
including also the different X-ray absorption edges to select the
structures involving particular atomic elements. This will allow the
determination of chemical/magnetic ordering of specific atoms belonging
to the crystal lattice. The introduction of this technique will therefore
be useful in special cases where microscopy techniques are not appropriate.

It is also important to understand why such diffuse satellites were
not previously observed in x-ray diffraction experiments \cite{key-9,key-10}:
there, one mainly observed Bragg peaks, which are sensitive to atomic
distances of individual terraces. The radiation is scattered by atoms
and variations in the terrace widths lead to an incoherent sum of
the x-ray waves from neighboring terraces. In our present reflectivity
method, one is not sensitive to the details of the crystal structure.
The x-ray wave does not lose coherency when scattered by each terrace
and the resulting interference pattern reveals the modulation of the
surface structure.

In summary, we demonstrate in this work that the resonant soft X-ray
diffuse reflectivity method is a powerful tool to investigate microscopic
structures such as the terrace-like structures of MnAs/GaAs(001) films.
We observed a clear long range periodic arrangement of stripe-like
micro-structures alternating between $\alpha$ and $\beta$ phases
during the phase coexistence, which takes place between 10 and 45
°C. The period of the modulation structure obtained is about 600 nm,
which remains constant with the temperature. The results are in good
agreements with those reported in similar MnAs films using different
experimental techniques.

The authors thank the personnel of the LURE laboratory for technical
support. R.M-P. and H.W.Jr. thank FAPESP, CNPq, FAPEMIG and Instituto
do Milênio (Nanociências) for financial support.


\begin{thebibliography}{11}
\bibitem{key-17}In this procedure we have subtracted the contribution of the specular
peak from the intensity of the first satelite peak, i.e., we have
substracted the nearly temperature independent component of the intensity
of the first satellite. In this case, $\frac{I\left(q_{x}=\frac{2\pi}{\ell}\right)}{I\left(q_{x}=0\right)}$
is aproximatelly given by the ratio of the second (p = 1) and first
terms of eq. 6, leading to eq. 7.
\end{thebibliography}
\end{document}